# Probing Spin Accumulation in Ni/Au/Ni Single-Electron Transistors with Efficient Spin Injection and Detection Electrodes

R. S. Liu[a,b], H. Pettersson[a,b]*, L. Michalak[c], C. M. Canali[c] and L. Samuelson[b]

[a] Center for Applied Mathematics and Physics, Halmstad University, Box 823, SE-301 18 Halmstad, Sweden

[b] Solid State Physics/ the Nanometer Structure Consortium, Lund University, Box 118, SE-22100 Lund, Sweden

[c] Dept of Chemistry and Biomedical Sciences, Kalmar University, 391 82 Kalmar, Sweden.

*Corresponding author. E-mail: Hakan.Pettersson@ide.hh.se

**Abstract**

We have investigated spin accumulation in Ni/Au/Ni single-electron transistors assembled by atomic force microscopy. The fabrication technique is unique in that unconventional hybrid devices can be realized with unprecedented control, including real-time tunable tunnel resistances. A grid of Au discs, 30 nm in diameter and 30 nm thick, is prepared on a $SiO_2$ surface by conventional e-beam writing. Subsequently, 30 nm thick ferromagnetic Ni source, drain and side-gate electrodes are formed in similar process steps. The width and length of the source and drain electrodes were different to exhibit different coercive switching fields. Tunnel barriers of NiO are realized by sequential Ar and $O_2$ plasma treatment. Using an atomic force microscope with specially designed software, a single non-magnetic Au nanodisc is positioned into the 25 nm gap between the source and drain electrodes. The resistance of the device is monitored in real-time while the Au disc is manipulated step-by-step with Ångstrom-level precision. Transport measurements in magnetic field at 1.7 K reveal no clear spin accumulation in the device, which can be attributed to fast spin relaxation in the Au disc. From numerical simulations using the rate-equation approach of orthodox Coulomb blockade theory, we can put an upper bound of a few ns on the spin-relaxation time for electrons in the Au disc. To confirm the magnetic switching characteristics and spin injection efficiency of the Ni electrodes, we fabricated a test structure consisting of a Ni/NiO/Ni magnetic tunnel junction with asymmetric dimensions of the electrodes similar to those of the SETs. Magnetoresistance measurements on the test device exhibited clear signs of magnetic reversal and a maximum TMR of 10%, from which we deduced a spin-polarization of about 22% in the Ni electrodes.

Electron tunneling through ferromagnetic junctions is of current interest due to expected applications in magnetic random access memories (MRAM), read/write heads in hard discs and in other spintronic devices[1]. Most of the experimental and theoretical work published up to now focus on tunnel magnetoresistance (TMR) behavior in simple planar junctions. In this context, TMR implies an increase in the junction resistance when the magnetic moments of the two leads change from parallel to antiparallel alignment. More recently, spin-dependent tunneling in more complex systems, e.g. ferromagnetic single-electron transistors (F-SETs), has become an attractive topic for both experimental and theoretical studies[2]. In these devices, novel phenomena are expected to occur due to the interplay between charging effects and spin-dependent transport. Indeed, several experiments on F-SETs, starting with the seminal work by Ono et al.[3,4], have demonstrated enhanced TMR[3,5], and magneto-Coulomb oscillations of the TMR[4,6] as a function of external magnetic field and bias voltage. On the other hand theoretical work[7-13], besides clarifying some of these earlier experimental observations, predicts that spin accumulation on the central island of an F-SET should manifest itself in a variety of effects showing up in the magnetoresistance properties. Spin accumulation, together with spin injection, is a key concept in spintronics, and refers to a non-equilibrium spin population created in confined structures by external magnetic fields or spin-polarized currents. So far experimental verification of spin accumulation in F-SETs has been elusive. Only very recently two experiments[14,15] have shown indirect evidence of its occurrence, albeit the interpretation of one of them[14] has been controversial[16].

In the following, we use the notation F/F/F and F/N/F for SETs with ferromagnetic leads and a ferromagnetic central island, or non-magnetic island, respectively. One fundamental difference between F/F/F and F/N/F SETs is the connection between spin accumulation and TMR. For F/F/F SETs a non-zero TMR can exist even in the case of vanishing spin accumulation on the central island[7]. In this case, discrete charging effects lead to TMR oscillations as a function of the bias voltage[7]. In contrast, for F/N/F SETs a net spin accumulation on the central island is necessary to observe a non-zero TMR at all. In this case the intrinsic spin-relaxation time on the central island is sufficiently long compared to the time interval between two successive tunneling events, so that a spin-polarized current generates a finite magnetic moment. The landmark of the occurrence of spin accumulation in both devices is a periodic sign change of the TMR as a function of the bias voltage, with dips directly related to the Fermi level splitting for electrons with different spin[9,11-13]. Such features were indeed observed in the experiment by Yakushiji et al.[15] on F-SETs with one magnetic lead and a central island consisting of a Co nanoparticle.

One important conclusion from Ref. 15 is that the crucial parameter that controls spin accumulation, namely the spin-relaxation time, $\tau_{RE}$, is apparently strongly enhanced in nanoparticles over bulk structures. For instance, $\tau_{RE}$ in Co nanoparticles is enhanced up to hundreds of nanoseconds in comparison with tens of picoseconds in ferromagnetic layers. Among the possible reasons for such long $\tau_{RE}$ is the suppression of the spin-orbit mediated spin-flip scattering caused by the discreteness of the nanoparticle energy levels,[17] or by the properties of the matrix in which the nanoparticle is embedded.[18] It is fair to say, however, that spin-relaxation mechanisms in metal nanoparticles are at present not well understood.

In this paper, we demonstrate a novel type of SET design suitable for studying spin accumulation in well-controlled, strongly confined nanoparticles. The nanoparticle is in our case attached to a SiO$_2$ surface and not embedded in a disturbing matrix. We have chosen to study nanoscaled Au discs since no conclusive reports on the dependence of confinement on the spin-relaxation mechanism have been reported. In addition, Au is inherently interesting due to its very strong spin-orbit interaction.

In the present work, F/N/F SETs are fabricated on top of a 100 nm thick SiO$_2$ layer grown on a Si substrate (shown in Figure 1 (A)). A grid of Au discs, 30 nm in diameter and 30 nm thick, is prepared by conventional electron-beam lithography followed by thermal evaporation and subsequent lift-off. Ferromagnetic Ni source and drain electrodes, 30 nm thick, 220 nm and 80 nm wide, 300 nm and 1.5 μm long, respectively, are formed in subsequent similar process steps together with a 30 nm thick Ni side-gate. Tunnel barriers of NiO are fabricated by sequential Ar plasma etching at a pressure of about $1.0 \times 10^{-4}$ mbar for 5 minutes and O$_2$ plasma etching at a pressure of about $1.0 \times 10^{-1}$ mbar for 3 minutes.

The width (80 nm) of the drain electrode is comparable to the size of single Ni domains[19,20], while the source electrode is considerably wider (220 nm) and shorter (300 nm). Because of this shape anisotropy, the two electrodes will undergo magnetic reversal at different magnetic fields, confirmed by performing micromagnetic simulations using the OOMMF code[21] which give the coercivity field of 40mT and 90mT for the source and drain electrodes, respectively. By sweeping the magnetic field it should thus be possible to switch from parallel to antiparallel alignment of the magnetic moments in the two electrodes. The magnetic switching behavior and spin-injection efficiency of the electrodes is investigated in a test structure shown in the inset of Fig. 1B. The test structure consists of a Ni/NiO/Ni magnetic tunnel junction with overlapping Ni electrodes of the same dimensions as those used for the SET. The junction was fabricated by first forming a Ni electrode using conventional methods. Following this, a sequence of plasma treatment steps forms the NiO tunnel barrier on top of the electrode. Subsequently, a second Ni top electrode is defined to overlap the bottom electrode with approximately 50nm using a high-precision alignment procedure. Magnetoresistance measurements clearly show a maximum TMR signal of about 10% while sweeping the magnetic field (Fig. 1B). This TMR signal provides strong support for that we can control the relative orientation of the magnetization of the two leads, and additionally, that the electrodes are efficient injectors of spin polarized current. Using Julliere´s model[22], TMR = $(R_{AP}-R_P)/R_P = 2P^2/(1-P^2)$, where $R_P$ and $R_{AP}$ are the resistances in parallel and anti-parallel magnetic configurations respectively, and $P$ is the spin-polarization, we deduce a spin-polarization of 22% in the Ni electrodes. This value is in good agreement with the tunneling spin-polarization measured by Tedrow and Meservey in planar tunnel junction experiments[23]. It is noted that the magnetic switching occurs for slightly lower fields than expected from the OOMMF simulations, a discrepancy probably due to exchange bias introduced by the antiferromagnetic NiO.

After mounting and bonding the real sample on a standard chip carrier, a Au nanodisc is positioned step-by-step with Angstrom precision into the 25nm gap between the drain and source electrodes using the AFM manipulation technique described in Ref. 24. The resistance of the device is monitored in real-time while the Au disc is manipulated, and the desired resistance (1 MΩ ~ 1 GΩ) for a good device can normally be obtained after merely a few attempts. For a given device, it is possible to tune the tunnel resistance by re-positioning the Au disc. A fairly high fabrication yield of typically 10% is obtained. The present devices differ significantly from previously reported AFM-assembled devices in that the electrodes are ferromagnetic with plasma-processed NiO tunnel barriers, and that only a single disc is used to bridge the gap between the source and drain electrodes. These substantial developments have resulted in well-controlled ferromagnetic SETs with long-term stability and a decreased fabrication complexity[25]. After the fabrication is completed, extensive conductance measurements are performed at 4.2K in a liquid helium dewar (schematic circuit diagram is shown in Fig. 1(C)). Following these measurements, the sample is transferred to a cryostat housing a 6T superconducting magnet where the magnetoresistance measurements were carried out at 1.7 K. The magnetic field is in the plane of the device with a tunable orientation with respect to the orientation of the electrodes.

Figure 2 shows a color-coded plot of the differential conductance *dI/dV* as a function of drain-source voltage Vd and gate voltage Vg. The dark areas correspond to Coulomb blockade regimes. The gate capacitance $C_G$ is determined from the spacing between neighboring degeneracy points at Vd=0 where *dI/dV* is non-zero, resulting in $C_G$ = e/(200mV) = 0.80 aF. The asymmetry observed in the Coulomb diamonds with respect to $V_G$ = 0 reflects the presence of a non-zero background charge. The drain-source threshold voltage required for tunneling of one electron through the device is given by $V_{th}$ = e/$C_\Sigma$ and amounts to about 2.5 meV. From this we estimate a total capacitance $C_\Sigma = C_S + C_D + C_G$ of 64 aF, and a corresponding charging energy $E_C=e^2/2C_\Sigma$ of 1.25 meV. From the absence of any Coulomb staircase in the I-V characteristics, it is evident that the rates with which electrons tunnel through the source and drain junctions are similar, hence $\Gamma_S = \Gamma_D$. The tunnel junctions can be regarded as plate capacitors with C = $\varepsilon_r \varepsilon_0 A/r$, where A is the area and r is the thickness of the tunnel barrier. Assuming similar capacitances for the source and drain C = $C_D$ = $C_S$ = 31.6 aF, $\varepsilon_{NiO} = 10.31$ and r = 1nm (estimated), we obtain an effective tunnel junction area of 346 nm$^2$ corresponding to 30 nm (height) × 11.5 nm (width).

Figure 3 shows a color coded plot of *dI/dV* as a function of the drain-source voltage Vsd and magnetic field B at $V_G$=0. The in-plane magnetic field, swept from -0.4T to 0.4T, was applied parallel to the drain-source electrodes. Evidently, no clear signs of TMR were observed when sweeping the magnetic field. In fact, an almost constant value of *dI/dV* was observed at the Coulomb blockade threshold voltage (white narrow strip), indicating the absence of spin accumulation. The same results were obtained when applying the magnetic field perpendicular to the source-drain electrodes. We have carried out magnetotransport measurements on 4 samples, and none of them gave any clear TMR signal. Clearly, some obvious causes for the lack of TMR, such as a poor quality of the magnetic tunneling junctions and poor spin injection efficiency of the Ni leads, can be ruled out from the observed 10% TMR of the test structure discussed above. To estimate the noise level of our device, we have calculated TMR values defined as $TMR = (I_{B_1} - I_{B_2})/I_{B_2}$ as a function of the bias voltage. Here $B_1$ was chosen to be -0.200 T in order to facilitate parallel magnetization of the electrodes, while different values of $B_2$ was selected in the interval -50 mT to 50 mT since the relative magnetization is expected to reverse in this region (see Fig. 1B). The data looks similar for different $B_2$ values and we plot a typical TMR curve for $B_2$ = 0 T in Fig.4.

An estimate of the expected magnetoresistance ratio at large bias is given by the well known expression $TMR = 2\frac{\tau_{RE}|P_1 P_2|}{e^2 \rho v (R_S + R_D)}$, where $\rho$ is the density of states in the Au disc, $v$ is the discs' volume, and $P_1$, $P_2$ denote the conduction electron spin-polarization of the source and drain electrodes respectively[11]. In our case the spin-polarization of the Ni electrodes is 22%. The denominator $e^2 \rho v (R_S + R_D)$ can be identified as an effective dwell time $\tau_{dwell}$. The spin-relaxation time $\tau_{RE}$, which is the most crucial parameter of the TMR effect, depends on the island material and can also be considerably affected by the small size of the nanoparticle. In order to get a more accurate quantitative measure of the expected TMR, and its dependence on the bias voltage, we have performed detailed numerical simulations using the rate-equation approach of orthodox Coulomb blockade theory. In dealing with F-SETs with noncollinear configurations, it is often necessary to use a more sophisticated theoretical treatment in the spin transport[26-29]. In our device only the external electrodes are magnetic and most likely their magnetization is parallel or antiparallel to the external magnetic field. Even in the case in which a certain degree of noncollinearity is present in the device, we believe that its effect is much more subtle than what we are after. It is plausible that a noncollinear configuration can to some extent modulate the TMR as a function of external field, but it will not completely suppress spin accumulation. In the following we thus consider only collinear configurations.

The parameters of the model, namely tunnel resistances and capacitances, are adjusted to give the best fit of the experimental tunnel current and threshold voltage of the Coulomb blockade region. The capacitances are taken to be $C_S$ = 24 aF, $C_D$ = 20 aF and $C_G$ = 0.8 aF. The tunnel resistances are both $R_S = R_D$ = 0.34 MΩ. We have also introduced an offset charge $Q_0$ = -0.25e to correctly reproduce the asymmetry of the I-V curve around zero gate bias. In Fig. 4 we plot the theoretically predicted TMR at T=1.7 K as a function of the bias voltage for a few values of the dimensionless relaxation time $\alpha = \tau_{RE}/\tau_{dwell}$, which enters the self-consistent equation for the splitting of the Fermi energy and thus controls the spin accumulation on the Au disc. Comparing the experimental and theoretical TMR in Fig.4, it is difficult to say if there is indeed any genuine TMR. The apparent sign change of the noisy experimental TMR at low biases is not reproduced in the simulations at this temperature and for this choice of SET parameters. In fact, from our calculations it is evident that devices with symmetrical tunnel junctions in general exhibit changes in the sign of TMR only at very low temperatures. In Fig 5 we show the calculated TMR at 0.1K which is the highest temperature for which negative dips in the TMR are observed outside the Coulomb blockade region. From Fig. 4, we furthermore note the presence of an experimental TMR value of about 1-2% at larger bias. This signal could obviously be interpreted as a genuine TMR signal, but it could also simply reflect a spurious charging effect. The discrepancy between the experimental results and numerical simulations hence makes the interpretation of the data in terms of TMR uncertain, leading us to settle with a determination of an upper bound for the spin relaxation time. We note that for $\alpha \approx 0.2$ the value of the theoretical TMR is of the order of 2%, which is approximately equal to the experimental TMR at large bias. With the choice of the tunnel resistances made above and using bulk density of states for Au, we estimate $\tau_{dwell} \approx 20$ ns. Using $\alpha = 0.2$ we obtain an upper bound for $\tau_{RE}$ of 4 ns in an Au island with dimensions of a few tens of nm, which is several order of magnitudes larger than the spin-relaxation time previously reported in thin Au films[30]. Given the uncertainty in the interpretation of our TMR signal, we emphasize that this estimate is only an upper bound for $\tau_{RE}$. Nevertheless this conclusion is still significant and complementary to other recent magnetotransport measurements on F-SETs with smaller Au nanoparticles, which report a $\tau_{RE}$ of the order of 1 ns[29, 31]. It should also be noted that our deduced spin-relaxation time is much larger than the spin-orbit scattering time $\tau_{SO}$ estimated in Au nanoparticles of comparable size by investigating the individual g-factors of the non-interacting electron states[32]. The time $\tau_{SO}$ represents merely an average strength of the spin-orbit interaction and does not correspond to any real relaxation process. The difference between these two times clearly shows that the strong spin-orbit interaction, certainly present in noble metals[33,34] and responsible for a very short $\tau_{SO}$ in nanoparticles, is only one variable controlling $\tau_{RE}$. The other crucial element leading to spin relaxation is the coupling of the electron spin to other degrees of freedom of the nanoparticle and the surrounding substrate, such as phonons and magnons. The microscopic mechanisms of this coupling, and their dependence on the nanoparticle size, are important problems in spintronics which need to be unraveled by further theoretical and experimental studies.

In conclusion, we have studied spin accumulation in a Au nanodisc by performing magnetotransport measurements on a novel type of SET design assembled with an AFM. Ferromagnetic Ni source and drain electrodes are realized using conventional e-beam writing. In the same process we also produce the Ni side-gate electrode. Different widths and lengths of the source and drain electrodes facilitate magnetic reversal of the two electrodes at different magnetic fields. Tunnel barriers of NiO are realized by sequential Ar and $O_2$ plasma treatment. Using an AFM with specially designed software, a single non-magnetic Au nanodisc is positioned into the 25 nm gap between the drain and source electrodes. From measurements of spin-polarized transport via the Au nanodisc we conclude that no clear signatures of TMR are seen, indicating that spin accumulation in the Au island is not occurring due to fast spin relaxation in the Au island. From comparison with results of theoretical modeling we deduce an upper bound of 4 ns for the spin-relaxation time in an Au island with dimensions of a few tens of nanometers. To investigate the switching characteristics and spin-injection efficiency of the Ni electrodes, reference Ni/NiO/Ni tunnel junctions with dimensions of the electrodes similar to those employed in the SETs were fabricated. A maximum 10% TMR was observed from which we deduce a conduction electron spin-polarization of about 22% in the Ni electrodes in good agreement with theory.

The authors thank A. Fuhrer and D. Suyatin for help with setting up the experiments and for fruitful discussions. The authors furthermore acknowledge financial support from Halmstad University, the Faculty of Natural Sciences at Kalmar University, the Swedish Research Council under Grant No: 621-2004-4439, the Swedish National Board for Industrial and Technological Development, the Office of Naval Research, and the Swedish Foundation for Strategic Research.

**Reference:**

[1] *Spin Electronics,* edited by M. Ziese and M.J. Thornton (Springer, Berlin Heidelberg, 2001).
[2] *Spin dependent transport in nanostructures*, edited by S. Maekawa and T. Shinjo (Taylor and Francis, London, 2002).
[3] K. Ono, H. Shimada, and Y. Ootuka, J. Phys. Soc. Jpn. **66**, 1261 (1997).
[4] K. Ono, H. Shimada, and Y. Ootuka, J. Phys. Soc. Jpn. **67**, 2852 (1998).
[5] K. Yakushiji *et al.*, Appl. Phys. Lett. **78**, 515 (2001).
[6] K. Yakushiji *et al.*, J. Appl. Phys. **91**, 7038 (2002).
[7] J. Barnas and A. Fert, Phys. Rev. Lett. **80**, 1058 (1998).
[8] S. Takahashi and S. Maekawa, Phys. Rev. Lett. **80**, 1758 (1998).
[9] J. Barnas and A. Fert, Europhys. Lett. **44**, 85 (1998).
[10] A. Brataas, Y. V. Nazarov, J. Inoue and G. E. W. Bauer, Phys. Rev. B **59**, 93(1999).
[11] A. N. Korotkov and V. I. Safarov, Phys. Rev. B **59**, 89 (1999).
[12] H. Imamura, S. Takahashi and S. Maekawa, Phys. Rev. B **59**, 6017(1999).
[13] I. Weymann and J. Barnas, Phys. Status Solidi B **236**, 651 (2003).
[14] C.D. Chen, Watson Kuo, D.S. Chung, J.H. Shyu, and C.S. Wu, Phys. Rev. Lett. **88**, 047004 (2002).
[15] K. Yakushiji *et al.*, Nature Materials **4**, 57 (2005).
[16] J. Johansson, M. Urech, D.B. Haviland, and V. Korenivski, Phys. Rev. Lett. **91**, 149701 (2003); C.D. Chen *et al., ibid.* **91**, 149701 (2003).
[17] G.G. Khalliulin and M.G. Khusainov Sov. Phys. JETP **67**, 524 (1988).
[18] G. Mitrikas, C.C. Trapalis, and G. Kordas, J. Chem. Phys. **111**, 8098 (1999)
[19] Y. Jaccard, Ph. Guittienne, D. Kelly, J.-E. Wegrowe, and J. -Ph. Ansermet, Phys. Rev. B **62**, 1141(2000).
[20] S.Y. Chou, P. R. Krauss, and L. Kong, J. Appl. Phys. **79**, 6101(1996).
[21] OOMMF is Object Oriented Micromagnetic Framework, a micromagnetic simulation code available free from NIST at http://math.nist.gov/oommf/.
[22] M. Julliere, Phy. Lett. A 54, 225 (1975).
[23] R. Meservey and P.M. Tedrow, Phys. Rep. **238**, 173 (1994).
[24] T. Junno, S.-B. Carlsson, Hongqi Xu, L. Montelius, and L. Samuelson, Appl. Phys. Lett. **72,** 548 (1998).
[25] R.S. Liu, D. Suyatin, H. Pettersson and L. Samuelson, accepted by Nanotechnology 2006.
[26] J. König and J. Martinek, Phys. Rev. Lett. **90,** 166602 (2003).
[27] M. Braun, J. König and J. Martinek et al., Phys. Rev. B **70,** 195345 (2004).
[28] S. Braig and P W. Brouwer, Phys. Rev. B, **71**, 195324 (2005).
[29] W. Wetzels and G E. W. Bauer and M. Grifoni, Phys. Rev. B. **72**, 020407 (2005).
[30] A.Y. Elezzabi, M.R. Freeman and M. Johnson, Phys. Rev. Lett. **77**, 3220 (1996).
[31] A. Bernand-Mantel, P Seneor, N.Lidgi, L. Calvet, V. Cros, K. Bouzehouane, S. Fusil, C. Deranlot, A.Vaures, F. Petroff and A. Fert, "*Spin dependent Coulomb Blockade in metallic nanoclusters*", poster presented at the conference: "Spin-dependent transport through nanostructures -- Spintronics '05, Poznan, September 2005.
[32] J.R. Petta and D.C. Ralph, Phys. Rev. Lett. **87**, 266801 (2001).
[33] Y. Yafet, Phys. Rev. **85**, 478 (1952).
[34] R.J. Elliott, Phys. Rev. **96**, 266 (1954).

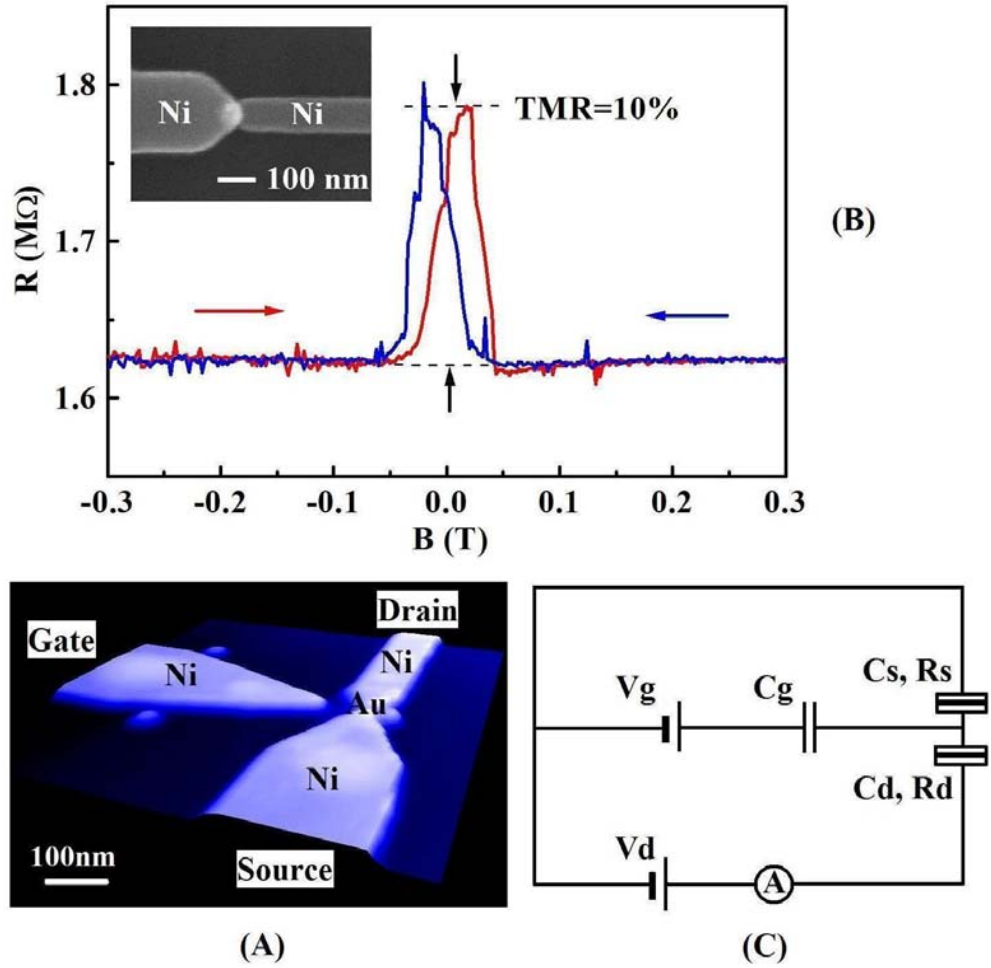

Fig. 1. (A) Atomic force micrograph of the ferromagnetic SET studied in the present work. The device is fabricated on top of a 100nm thick $SiO_2$ layer. (B) Switching behavior of a tunnel junction between two Ni electrodes separated by NiO as a function of magnetic field at 1.7K. The inset shows a scanning electron micrograph of the device. (C) Schematic circuit diagram of the SET.

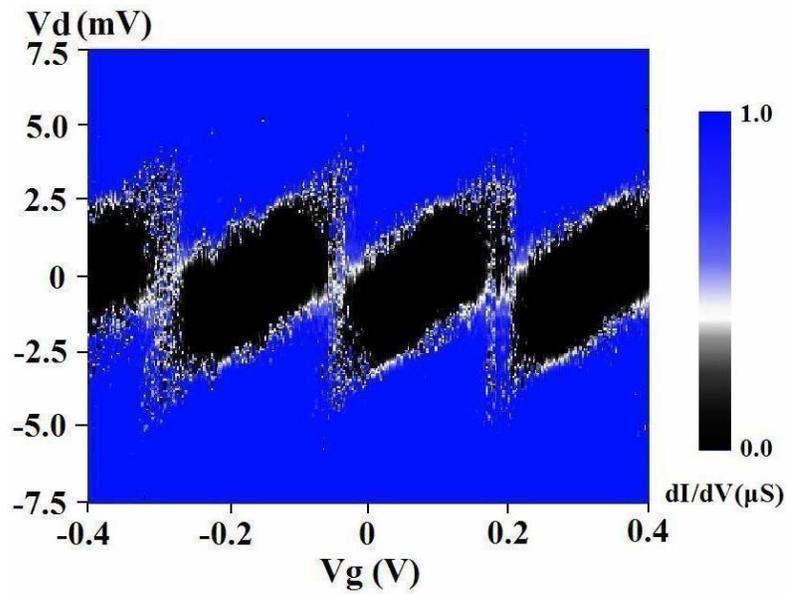

Fig. 2. Color coded plot of the differential conductance *dI/dV* as a function of drain-source voltage Vd and gate voltage Vg obtained at 4.2K. The dark areas correspond to Coulomb blockade regions

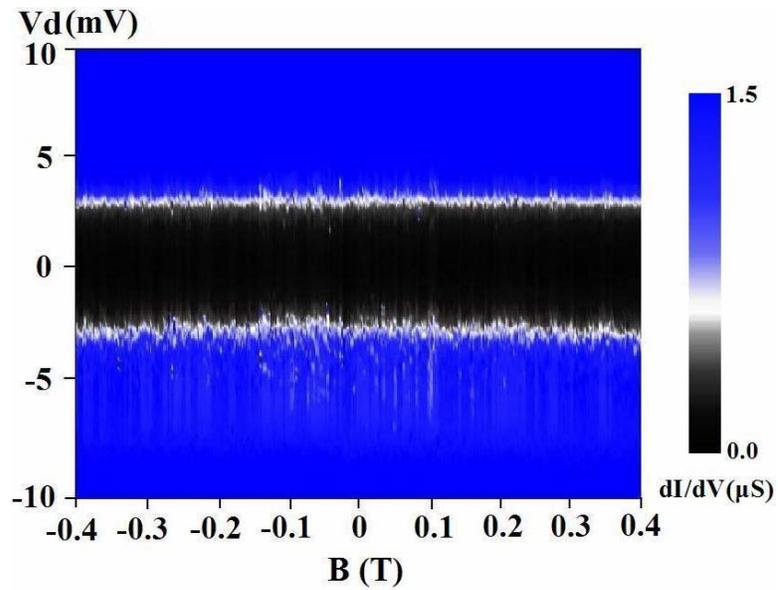

Fig.3. Color coded plot of *dI/dV* obtained at 1.7K as a function of the drain-source voltage Vd and magnetic field B at $V_G = 0$.

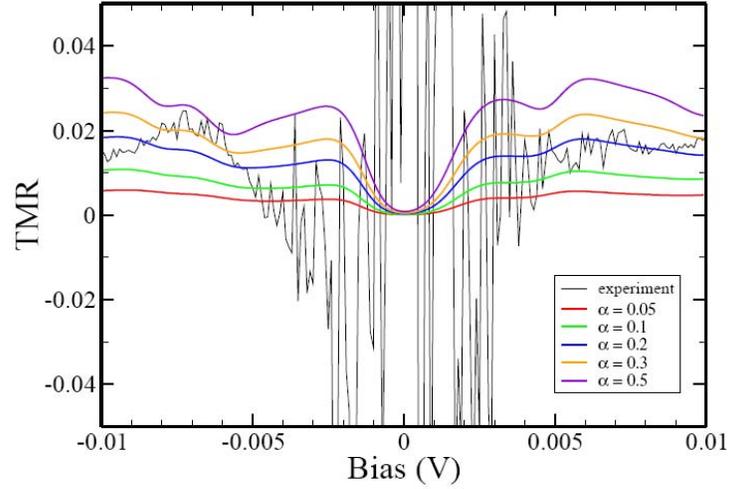

Fig. 4. TMR as a function of the drain-source bias at T = 1.7 K. The solid curve is the experimentally obtained TMR signal. The other curves are theoretical TMR signals calculated within the orthodox theory for several values of the dimensionless spin relaxation time α.

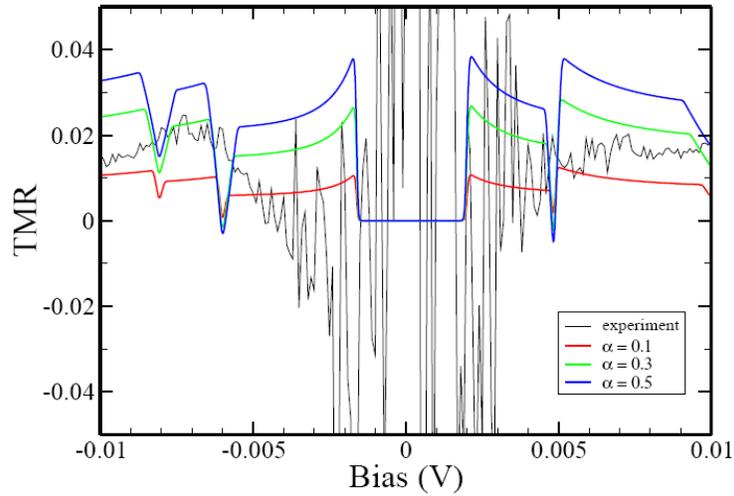

Fig. 5. TMR as a function of the drain-source bias. A temperature of 0.1 K was chosen in the simulations. The solid curve is the experimentally obtained TMR signal, recorded at 1.7K, adopted from Fig. 4.